\documentclass{ws-ijmpcs}
\usepackage{color}
\newcommand{\beq}{\begin{equation}}
\newcommand{\eeq}{\end{equation}}
\newcommand{\bea}{\begin{eqnarray}}
\newcommand{\ena}{\end{eqnarray}}

\newcommand{\ie}{{\it i.e.}}
\newcommand{\Sa}{\mbox{$\sigma_{\rm an}$}}
\definecolor{cyan}{cmyk}{1.,0.,0.,0.2}
\definecolor{vert}{cmyk}{0.5,0.,0.5,0.5}
\definecolor{magenta}{cmyk}{0.,1.,0.,0.1}
\definecolor{verdatre}{cmyk}{0.5,0.,0.5,0.5}
\definecolor{vert_clair}{cmyk}{0.5,0.,0.5,0.2}
\definecolor{yellow}{cmyk}{0.,0.,1.,0.0}
\definecolor{yellow_1}{cmyk}{0.,0.,0.5,0.0}
\definecolor{rouge}{cmyk}{0.,0.4,0.6,0.0}
\definecolor{orange}{cmyk}{0.,0.5,0.5,0.05}
\definecolor{violet}{rgb}{0.5,0.,0.5}
\definecolor{darwin_box}{rgb}{0.988,0.878,0.77}
\definecolor{darwin_text}{rgb}{0.1,0.07,0.02}
\definecolor{blue_mertsch}{rgb}{0.08,0.05,0.25}
\definecolor{blue_light_mertsch}{rgb}{0.7,0.67,0.77}
\begin{document}
\markboth{P. Salati}
{Dark Matter Annihilation in the Universe}

%
\catchline{}{}{}{}{}
%
\title{DARK MATTER ANNIHILATION IN THE UNIVERSE}

\author{PIERRE SALATI}

\address{LAPTh, Universit\'e de Savoie \& CNRS, 9 chemin de Bellevue, B.P.110\\
Annecy-le-Vieux, F-74941, France\\
pierre.salati@lapth.cnrs.fr}

\maketitle

\begin{history}
\received{12 February 2014}
\end{history}

\begin{abstract}
The astronomical dark matter is an essential component of the Universe and yet its nature is still unresolved.
It could be made of neutral and massive elementary particles which are their own antimatter partners. These
dark matter species undergo mutual annihilations whose effects are briefly reviewed in this article.
Dark matter annihilation plays a key role at early times as it sets the relic abundance of the particles once they
have decoupled from the primordial plasma. A weak annihilation cross section naturally leads to a cosmological
abundance in agreement with observations.
Dark matter species subsequently annihilate -- or decay -- during Big Bang nucleosynthesis and could play havoc
with the light element abundances unless they offer a possible solution to the $^{7}$Li problem.
They could also reionize the intergalactic medium after recombination and leave visible imprints in the cosmic
microwave background.
But one of the most exciting aspects of the question lies in the possibility to indirectly detect the dark matter
species through the rare antimatter particles -- antiprotons, positrons and antideuterons -- which they produce
as they currently annihilate inside the galactic halo.
Finally, the effects of dark matter annihilation on stars is discussed.
%
\end{abstract}

\ccode{PACS numbers: 95.35.+d, 98.80.Cq, 95.30.Cq, 26.35.+c, 95.85.Ry, 96.50.S-, 98.70.Sa, 97.10.Cv}

\vskip 0.1cm
\rightline{LAPTH-Conf-011/14
\footnote{Proceedings of 2nd International Workshop on Antimatter and Gravity WAG 2013 held in Bern.}
{\hspace{0.6cm}}}

\section{Dark Matter production through annihilation}

Large amounts of invisible matter in the Universe have been discovered in 1933
when the Swiss astronomer Fritz Zwicky measured~\cite{1933AcHPh...6..110Z}
the velocity dispersion of individual galaxies inside the Coma cluster. That
self-gravitating system contains thousands of objects and has quite certainly
virialized given its age and spherical shape. Because a steady state has been
reached, the cluster mass and size can be related to the velocity
dispersion of the galaxies which it shelters. Zwicky determined for the first
time the dynamical mass of the Coma cluster and obtained a value more
than a hundred times larger than the visible counterpart inferred from the luminosity
of galaxies. The astronomical dark matter puzzle originates from this measurement.
Since then, it has been continuously confirmed by an impressive series of refined
observations performed  with quite different methods.
The analysis of the X-ray emission from the hot gas filling
clusters allows to reconstruct their gravitational potential wells
and to infer their dynamical mass. The weak and strong lensing
of distant sources can also be used to derive the amount of material bending
the light trajectories. The dynamical to visible mass ratio is always very large.
Clusters can be pictured as icebergs with a tiny emerged visible part and a by
far dominant hidden component made of dark matter (DM).

The problem also exists on galactic scales as demonstrated by
Vera Rubin~\cite{1980ApJ...238..471R} and
Albert Bosma~\cite{1979A&A....79..281B} in 1979.
The rotation curves of spirals are determined with the help of the Doppler effect
through the 21 cm line emission from orbiting clouds of neutral atomic hydrogen HI.
The rotation speed of the disk does not exhibit the Keplerian decrease which would
be typical of a central and dominating mass. On the contrary, it is found to remain
flat far beyond the optical radius. The dynamical mass of spirals does not lie therefore
in their bulges but is spread over an extended halo of unseen material.

More recently, the measurements of the cosmic microwave background (CMB) anisotropies
by the Planck satellite in 2013 provide clear evidence~\cite{Ade:2013zuv} for a flat universe
whose density $\Omega_{\rm tot} = 1$ is equal to the closure value. A fluid with negative
pressure, dubbed dark energy or quintessence, coexists with non-relativistic matter. Dark
energy contributes a fraction $\Omega_{\Lambda} = 0.6825$ to the total mass budget
whereas matter amounts to $\Omega_{\rm M} = 0.3175$. Nucleons and electrons which make
up the so-called baryonic matter contribute only a small fraction $\Omega_{\rm B} = 0.049$
in agreement with primordial nucleosynthesis. A dark component with density
$\Omega_{\rm DM} = \Omega_{\rm M} - \Omega_{\rm B} = 0.2685$ appears on cosmological
scales with the surprising property of being made of an unknown form of matter.

Theoreticians have been imagining for the last three decades a plethora of candidates for this
astronomical dark matter. In spite of the proliferation of more or less exotic models,
the interest of the community has focused on the supersymmetric or Kaluza-Klein
extensions of the standard model of particle physics. These theories are based on a
new symmetry of Nature and naturally predict the existence of a weakly interacting
and massive particle -- dubbed WIMP -- whose mass lies in the GeV to
TeV range with typically weak interactions. This species is moreover stable because
of the conservation of the quantum number associated to the new symmetry. It is
electrically neutral and is its own anti-particle. A WIMP pair can annihilate to produce
standard particles like fermions or gauge bosons.
\beq
{\chi} + {\chi} \rightleftharpoons f + \bar{f} \, , \, W^{+} \! + W^{-} \, , \,
Z^{0} + Z^{0} \, , \, \cdots
\label{annihil_reac}
\eeq
Dark matter annihilation plays a crucial role in the early Universe as it provides
a natural mechanism through which WIMPs have been produced. Because
reaction~(\ref{annihil_reac}) is in equilibrium during the Big Bang,
DM species exist under the form of an ultra-relativistic radiation as long
as the temperature of the primordial plasma exceeds their mass. For a
10~GeV particle, this happens before an age of a few ns.
As soon as this condition stops to be fulfilled, WIMPs annihilate without being
produced back from lighter particles. Their density drops significantly until they
are so diluted that they stop interacting with each other. This chemical quenching
leaves a steady population of particles whose density decreases as space expands.
Because they are stable, WIMPs contribute today to the mass budget of the Universe.
Actually, their cosmological relic abundance is found~\cite{PhysRevLett.39.165}
to depend only on the annihilation cross section with
\beq
\Omega_{\chi} h^{2} \, = \,
{\displaystyle \frac{3 \times 10^{-27} \; {\rm cm^{3}} \; {\rm s^{-1}}}{< \! \Sa v \! >}}
\;\; ,
\eeq
where $h$ is the Hubble constant. With a cross section
$< \! \Sa v \! > \sim 3 \times 10^{-26} \; {\rm cm^{3}} \; {\rm s^{-1}}$
typical of weak interactions, the cosmological abundance $\Omega_{\chi}$ falls
naturally close to the observed value of $\Omega_{\rm DM} \simeq 0.27$. This coincidence
is called the WIMP miracle and is the reason why this type of particle is considered as
the favoured candidate to the astronomical dark matter. 

\section{Dark Matter Annihilation and Big Bang Nucleosynthesis}

Big Bang nucleosynthesis (BBN) is in remarkable agreement with observations of light
element abundances although some tension exists for $^{7}$Li. That species is essentially
produced through {$^{4}$He + $^{3}$He $\to$ $^{7}$Be + $\gamma$}
with subsequent electron capture of $^{7}$Be into $^{7}$Li. The BBN theory predicts
an abundance of
${\rm {^{7}Li}/{H}} = 5.24^{+0.71}_{-0.67} \times 10^{-10}$
whereas a plateau -- the so-called Spite plateau -- is measured at a level of
${\rm {^{7}Li}/{H}} = 1 - 2 \times 10^{-10}$ regardless of the metallicity.
Moreover, in standard BBN, the {$^{6}$Li} isotope is formed in the reaction
{$^{4}$He + D $\to$ $^{6}$Li + $\gamma$} which is extremely inefficient, hence
a theoretical value of ${\rm {^{6}Li}/{H}} \sim 10^{-14}$. Significant traces of that
element have nevertheless been found in the low-metallicity halo star HD84937 for
which ${\rm {^{6}Li}/{^{7}Li} = 0.052 \pm 0.019}$.

Primordial nucleosynthesis can be significantly perturbed by the injection of energetic and
non-thermal particles produced by the residual annihilation or decay of WIMPs.
DM species could play havoc with the synthesis of light elements in many
ways~\cite{Jedamzik:2009uy}.
To commence, electromagnetic products -- electrons, positrons and photons -- generate
showers and can lead to the photodissociation of nuclei.
Photons injected in the primordial plasma generate electromagnetic showers
as long as their energy $E_{\gamma}$ exceeds the threshold energy
$E_{C} \simeq {m_{e}^{2}}/{22 \, T}$ for pair production on the CMB.
That process is dominant because the baryon-to-photon ratio is so small.
Alternatively, shower photons with $E_{\gamma} \leq E_{C}$ can pair produce
on protons and $^{4}$He nuclei. They also Compton scatter off plasma electrons.
They have finally a small chance to photodisintegrate D below 3 keV and later on {$^{4}$He}
below 0.3 keV. Nuclei can be destroyed as soon as the cut-off energy $E_{C}$ exceeds their
photodissociation threshold. The larger the latter, the smaller the temperature at which
destruction starts to take place. The dissociation of {$^{4}$He} produces {D} and {$^{3}$He}
and leads to an abnormally large {$^{3}$He}/{D} ratio.

WIMP annihilation or decay can also produce hadronic particles.
Injection of $\pi^{\pm}$ induces charge exchange reactions {$\pi^{-}$ + p $\to$ $\pi^{0}$ + n}
for temperatures between 1~MeV and 300 keV. Creation of extra neutrons after the
neutron-to-proton ratio {n}/{p} freeze-out implies an increase of the helium mass fraction
$Y_{\rm p}$.
The same effect occurs if antinucleons are injected in the primordial plasma since they
preferentially annihilate on protons, thereby raising the effective {n}/{p} ratio.
Any extra neutrons injected at $T \sim 40$ keV can lead to an important depletion of $^{7}$Be.
Depending on its magnitude, this mechanism could solve the lithium problem.
Finally, at lower temperatures, energetic neutrons and protons can destroy $^{4}$He through
{n + $^{4}$He $\to$ $^{3}$H ($^{3}$He) + p (n) + n + ($\pi$'s)} or
{n + $^{4}$He $\to$ D + p + 2n + ($\pi$'s)}. This can result in the overproduction of {D}
but may also lead to a much larger $^{6}$Li abundance than predicted by standard BBN.

Non-thermal BBN provides a framework to solve the $^{7}$Li problem. If a small admixture
of neutrons is injected by WIMP annihilation or decay during or just after the synthesis of
$^{7}$Be, {\ie}, for temperatures between 60 and 30~keV, this element is converted into $^{7}$Li
via the neutron capture {n + $^{7}$Be $\to$ p + $^{7}$Li} and is efficiently destroyed by protons
through {p + $^{7}$Li $\to$ $^{4}$He + $^{4}$He}. This sequence of reactions leads to the
depletion of $^{7}$Be which is no longer transmuted into $^{7}$Li at later times by electron
capture, hence a residual $^{7}$Li abundance closer to the Spite plateau. That scenario is
nevertheless constrained by the requirement that extra neutrons should not overproduce {D}.

Energetic {$^{3}$He} and {$^{3}$H} may be produced via the spallation (hadronic) or
photodissociation (electromagnetic) reactions mentioned above. These nuclei collide on
$^{4}$He to produce {$^{6}$Li} through the endothermic reactions
{$^{3}$H  + $^{4}$He $\to$ $^{6}$Li + n - 4.78 MeV} and
{$^{3}$He + $^{4}$He $\to$ $^{6}$Li + p - 4.02 MeV}.
For projectiles with energy $\sim$ 10~MeV, the cross sections for these reactions are
$10^{7}$ times larger than the cross section of the
{$^{4}$He + D $\to$ $^{6}$Li + $\gamma$} standard $^{6}$Li production mechanism.
This opens the possibility for a much larger amount of $^{6}$Li than previously anticipated
and could explain the observations of the star HD84937. Conversely, $^{6}$Li is a sensitive
probe of DM annihilation. The cross section for WIMPs annihilating into ${q \bar{q}}$
pairs is constrained~\cite{Jedamzik:2009uy} to be smaller than
$8 \times 10^{-25}$ cm$^{3}$ s$^{-1}$ $({m_{\chi}}/{\rm 100 \; GeV})^{2}$
-- where ${m_{\chi}}$ is the WIMP mass --
under the penalty of yielding a {$^{6}$Li}/{$^{7}$Li} ratio in excess of 0.1.

\section{Cosmic Microwave Background Constraints}

Recombination of the primordial plasma takes place at a redshift ${\rm z} \simeq 1090$.
The electron fraction drops at a level of $10^{-4}$ and the intergalactic medium (IGM)
becomes neutral. Residual WIMP annihilation taking place at that time can perturb the
ionization history of the IGM and leave visible imprints on the CMB. Measurements
of the CMB temperature and polarization angular power spectra set constraints on the
energy deposited in the IGM just after recombination, thereby probing DM annihilation
occuring during that epoch.

WIMPs annihilate to a wide range of particles. However, the heating and ionization of
the IGM result mainly from the injection of electrons, positrons and photons which trigger
electromagnetic showers and eventually thermalize with the ambient medium.
As shown in~\refcite{Slatyer:2009yq}, the cooling of photons injected at a redshift ${\rm z} = 1000$
proceeds through various reactions. As the photon energy increases, the dominant mechanisms
are photoionization of IGM atoms, Compton scattering off electrons, pair production on the
H/He gas, photon-photon scattering and pair production on the CMB. All these processes
have timescales much smaller than the Hubble time so that thermalization effectively occurs.

The CMB constraints on possible modifications of the IGM ionization history after
recombination translate~\cite{Galli:2009zc,Slatyer:2009yq} into an upper limit of
$\sim 4 \times 10^{-25}$ cm$^{3}$ s$^{-1}$ $({m_{\chi}}/{\rm 100 \; GeV})$ on
the cross section of WIMPs annihilating into electromagnetic species. This value compares
to the BBN bound mentioned previously in the case of a pure ${q \bar{q}}$ channel.
Notice that most of the WIMP models which account for the cosmic ray positron excess
discussed in the next section are about to be ruled out by this CMB limit. The Planck satellite
will soon be able to set a more stringent bound and could exclude all these models.

\section{Indirect Signatures of Dark Matter Species}

Should WIMPs pervade the halo of the Milky Way, their mutual annihilations
would yield several indirect signatures. Although DM annihilation proceeds now at a
very small pace, it may nevertheless lead to the production of high-energy photons
and of rare antimatter particles such as antiprotons $\bar{p}$, positrons $e^{+}$
or even antideuterons $ \bar{D}$ through the reaction
\beq
\chi + \chi \; \to \; q + \bar{q} \, , \, W^{+} + W^{-} \, , \, \ldots \; \to \;
\bar{p} , \bar{D} , e^{+} \, \gamma \; \& \; \nu \;\; .
\label{indirect_reac}
\eeq
Antimatter species are already manufactured by conventional astrophysical processes.
The dominant mechanism is the spallation of primary cosmic ray (CR) protons and
helium nuclei on the gas of the galactic plane. Positrons could also be accelerated by
nearby highly-magnetized neutron stars called pulsars.
The messengers of DM annihilation would generate distortions in the signals detected
at the Earth or would appear in the $\gamma$-ray sky as hot spots with no optical
counterpart -- see the review~\refcite{Lavalle:2012ef} for more details.

\begin{figure}[t!]
\centerline{\includegraphics[width=10.0cm]{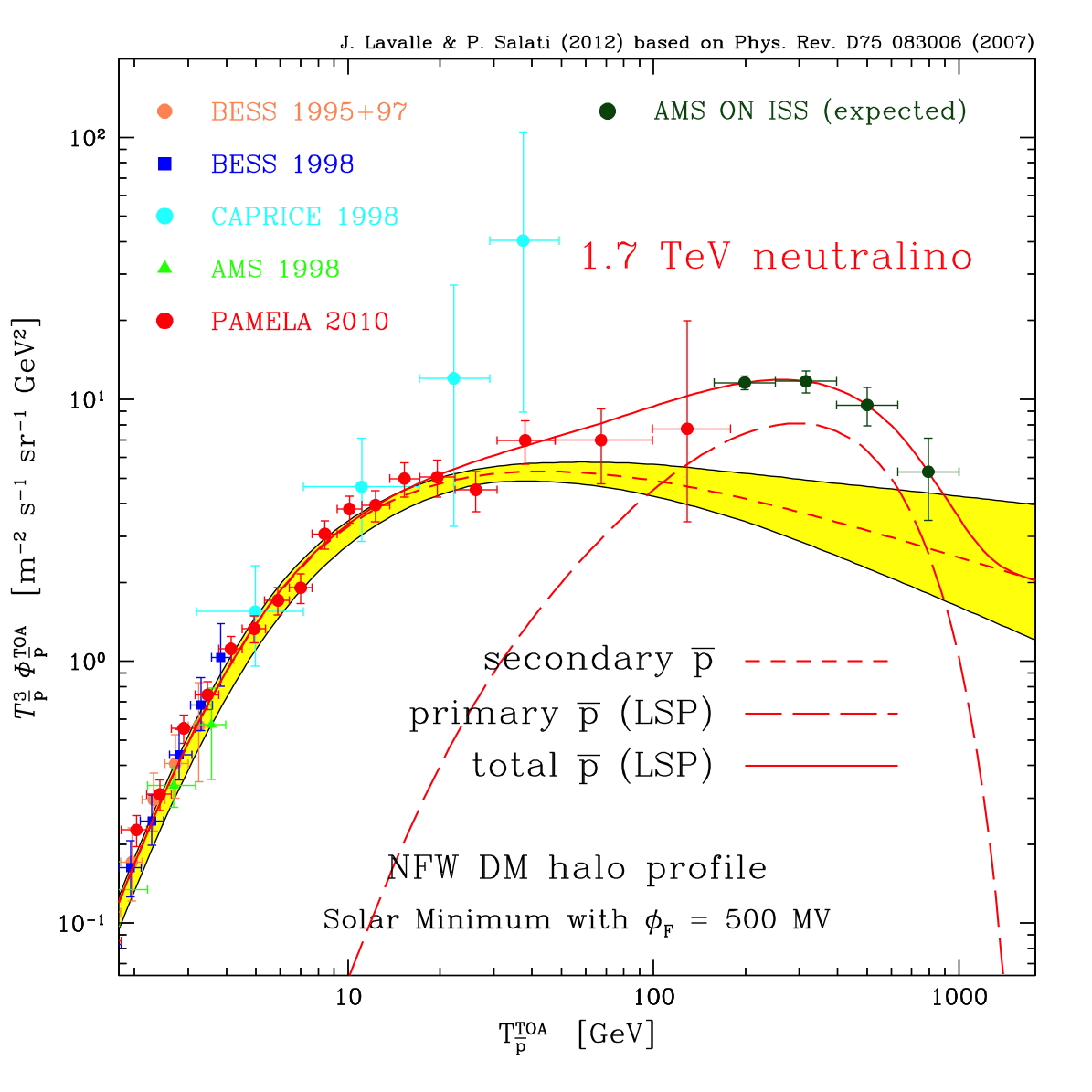}}
\vspace*{-8pt}
\caption{
The yellow band features the expected antiproton background for the full range of
diffusion parameters allowed by the B/C ratio. A heavy WIMP is also considered.
This DM species is almost a pure Wino and its annihilation cross section is
significantly enhanced today by non-perturbative, binding energy effects.
The corresponding primary (long dashed) and total (solid) fluxes have been
derived for a NFW halo profile and for the set of diffusion parameters that best fits
the B/C ratio. For illustration, a global boost factor of 2 has also been included in the
signal. The antiproton flux is compared to several measurements, whereas the
expected statistical error after 3 years of data sampling by AMS-02 is indicated.
}
\label{fig:heavy_wino}
\vskip -0.25cm
\end{figure}

Background antiprotons are produced inside the Galactic disk by the collisions undergone
by primary CR nuclei on H/He . Because they are not directly injected in the interstellar
medium (ISM) but are sourced by primary CR species, these astrophysical antiprotons are
dubbed secondaries.
In addition to this conventional mechanism, a primary component can be directly produced
by DM annihilation all over the Galaxy.
Irrespective of the production mechanism, antiprotons propagate inside the magnetic fields
of the Milky Way like any charged CR particle. Their transport may be modeled as a
diffusion process taking place inside a confinement domain called the magnetic halo. The
value of the diffusion coefficient, its dependence on the CR rigidity and the strength of
the convective wind that blows the particles away from the disk are unknown parameters
which can be constrained from the typical secondary-to-primary B/C ratio.
Although these constraints are far from being stringent, the flux of background antiprotons
is fairly well determined~\cite{Bringmann:2006im}
as featured by the yellow band of Fig.~\ref{fig:heavy_wino}.
Boron nuclei and secondary antiprotons are produced within the disk from primaries
interacting with the ISM. The similarity of the production mechanisms translates into a tight
relation between the B/C and ${\rm \bar{p}}/{\rm p}$ ratios, hence a good precision on
the antiproton background.
This is not the case for antiprotons produced by DM annihilation. This process takes place
all over the Galaxy and is not confined in its disk. The corresponding flux depends sensitively
on the thickness of the CR magnetic halo and is subject to an uncertainty which can reach two
orders of magnitude.
The DM halo profile is also unknown but the resulting error amounts to a factor $\sim$ 2
once the DM density $\rho_{\odot}$ in the solar neighborhood is fixed.
%
%

The announcement of a positron excess by the PAMELA collaboration~\cite{Adriani:2008zr}
in 2008 has triggered a lot of excitement. This excess was actually considered as the first hint of the
presence of WIMPs in the Milky Way halo. Five years later, the dust has settled down and most of
the initial hope has faded away.
As is clear from the AMS-02 measurements~\cite{PhysRevLett.110.141102}, the anomaly extends up
to 300~GeV and points towards a massive species, in good agreement with theoretical expectations.
Remember that
if WIMPs are thermally produced during the Big Bang, their relic abundance matches the Planck
value~\cite{Ade:2013zuv} of $\Omega_{\chi} \simeq 0.27$ provided that their annihilation cross section,
at the time of decoupling, is equal to
$< \! \Sa v \! > \sim 3 \times 10^{-26}$ cm$^{3}$ s$^{-1}$. Moreover, high-energy positrons cannot
diffuse on long distances, and those detected at the Earth must have been produced locally, hence a
DM density of $\rho_{\odot} = 0.3$ GeV cm$^{-3}$. Baring in mind these benchmark values,
one finds that the signal at the Earth is way too small to account for the observed excess. For a
WIMP mass $m_{\chi}$ of 1~TeV, the positron production rate needs to be enhanced by a factor of
$\sim 10^{3}$ to match the measurements.
Another difficulty lies in the absence of an antiproton excess. DM particles cannot couple to quarks
under the penalty of overproducing antiprotons~\cite{Cirelli:2008pk,Donato:2008jk}. Therefore,
besides an abnormally large annihilation rate today, WIMPs should preferentially annihilate into
charged leptons, a feature which is unusual in supersymmetry.
The positron excess is presumably produced by nearby pulsars.


\section{Dark Matter Annihilation and Stellar Evolution}

DM particles could play a role in stellar evolution. They may be captured by the stars which
they happen to cross as they wander in the Galaxy. WIMPs have a non-vanishing chance to
collide on a nucleus inside stellar interiors and to lose enough energy to become gravitationally
bound. A population of DM species builds up as time passes on and concentrates at the stellar
cores.
Because WIMPs interact weakly with their surroundings, they can transport heat on large
distances with a better efficiency than radiative diffusion. The central temperature gradient
is lowered. This scenario has actually been proposed~\cite{1985ApJ...294..663S} in 1985 to solve
the solar neutrino puzzle. But DM annihilation is a counteracting process that limits the
growth of the WIMP density. Once taken into account, the WIMP explanation of the
solar neutrino puzzle is no longer tenable.

In 1989, the focus was on DM annihilation inside stars. If DM haloes form from an
initial spherical configuration, they collapse into a central spike where the
WIMP density can be extremely large. This could be the case at the Galactic center.
A star floating in this environment would capture WIMPs at such a pace that DM
annihilation inside the object would provide enough energy to perturb the stellar
evolution. As shown in~\refcite{1989ApJ...338...24S}, main sequence stars would become red
giants and would be shifted in the H-R diagram towards the Hayashi track.
But these effects occur for large WIMP capture rates and require a WIMP-nucleus
scattering cross section of order 1 picobarn. Direct detection experiments have set
very stringent upper limits on that cross section and the scenario faded away.

The idea that WIMPs can significantly alter stellar evolution was revived in 2007.
During the dark age, at a redshift ${\rm z}$ between 10 and 50, DM is much denser
than today, especially at the centers of DM proto-haloes. These substructures act as
potential wells inside which baryons collapse as soon as molecular hydrogen becomes
sufficiently abundant to allow efficient gas cooling. The first generation of stars, {\ie},
the so-called population III, starts shining. The scenario proposed
by~\refcite{Spolyar:2007qv}
is based on the backreaction of the collapsing gas cloud on the surrounding DM distribution.
Adiabatic contraction drags WIMPs inwards as the gas concentrates and heats up.
Eventually, a dark star is born powered by the annihilation of the DM species which
it contains. The object is much cooler and more extended than standard Pop III stars.
The surface temperature does not exceed $2 \times 10^{4}$ K, hence a very weak UV emission.
The surrounding gas is not ionized and can cool down, thereby falling on the dark star
which it feeds with more gas and more WIMPs. The stellar mass increases up to
$\sim 800$ M$_{\odot}$ or even beyond~\cite{Spolyar:2009nt}.
Eventually, DM annihilation is no
longer able to power the star which contracts into a very massive object powered
by nuclear fusion. Its mass is so large that it may end its life as a hypernova.
Dark stars could be detected in the infrared with the James Webb space telescope
pointing through the central regions of foreground galaxy clusters acting as
gravitational lenses.

\section*{Acknowledgments}

P.S. would like to thank the organizers of the WAG 2013 meeting for their warm hospitality
and the friendly and inspiring atmosphere of the conference. This work has been supported
by Institut universitaire de France.

\bibliographystyle{ws-ijmpcs}
\bibliography{wag_biblio}
\end{document}